\documentclass[english]{article}
\usepackage[T1]{fontenc}
\usepackage[latin9]{inputenc}
\usepackage{geometry}
\usepackage{float}
\usepackage{url}
\usepackage{amsthm}
\usepackage{amsmath}
\usepackage{amssymb}
\usepackage{graphicx}
\usepackage{mathrsfs,url}

\makeatletter
\usepackage{enumitem}		% customizable list environments
\theoremstyle{plain}
\newtheorem{thm}{\protect\theoremname}[section]
  \theoremstyle{definition}
  \newtheorem{defn}[thm]{\protect\definitionname}
  \theoremstyle{plain}
  \newtheorem{prop}[thm]{\protect\propositionname}
  \theoremstyle{plain}
  \newtheorem{lem}[thm]{\protect\lemmaname}
  \theoremstyle{plain}
  \newtheorem{cor}[thm]{\protect\corollaryname}
  \theoremstyle{remark}
  \newtheorem{rem}[thm]{\protect\remarkname}

\usepackage{ifpdf} % part of the hyperref bundle
\ifpdf % if pdflatex is used

 \IfFileExists{lmodern.sty}{\usepackage{lmodern}}{}

\fi % end if pdflatex is used

\usepackage[all]{xy}

\usepackage{enumitem}

\usepackage{indentfirst}

\usepackage{tocbibind}

\date{}

\makeatother

\usepackage{babel}
  \providecommand{\corollaryname}{Corollary}
  \providecommand{\definitionname}{Definition}
  \providecommand{\lemmaname}{Lemma}
  \providecommand{\propositionname}{Proposition}
  \providecommand{\remarkname}{Remark}
\providecommand{\theoremname}{Theorem}

\begin{document}

\title{\textbf{Generalized solution for the Herman Protocol Conjecture}}

\author{Endre Cs\'oka%
\thanks{Alfr\'ed R\'enyi Institute of Mathematics. Supported by the NRDI grant KKP~138270 and by the European Research Council (grant agreement no.\textasciitilde{}306493).%
}, Szabolcs M\'esz\'aros%
\thanks{Department of Mathematics, Central European University%
}, Andr\'as Pongr\'acz%
\thanks{Alfr\'ed R\'enyi Institute of Mathematics. Supported by the NRDI grant KKP~138270.%
}}

\maketitle
\begin{abstract}
The Herman Protocol Conjecture states that the expected time $\mathbb{E}(\mathbf{T})$ of Herman's self-stabilizing algorithm in a system consisting of $N$ identical processes organized in a ring holding several tokens is at most $\frac{4}{27}N^{2}$. 
We prove the conjecture in its standard unbiased and also in a biased form for discrete processes, and extend the result to further variants where the tokens move via certain L\'evy processes.  
Moreover, we derive a bound on the expected value of $\mathbb{E}(\alpha^{\mathbf{T}})$ for all $1\leq \alpha\leq (1-\varepsilon)^{-1}$ with a specific $\varepsilon>0$. 
Subject to the correctness of an optimization result that can be demonstrated empirically, all these estimations attain their maximum on the initial state with three tokens distributed equidistantly on the ring of $N$ processes.
Such a relation is the symptom of the fact that both $\mathbb{E}(\mathbf{T})$ and $\mathbb{E}(\alpha^{\mathbf{T}})$ are weighted sums of the probabilities $\mathbb{P}(\mathbf{T}\geq t)$. 
\end{abstract}

\bigskip


\section{Introduction}

The simplified setup of Herman's self-stabilizing algorithm consists
of a directed circular graph of $N$ elements and $K$ tokens put
on $K$ different nodes of the graph. The vertices represent identical
processes connected along the edges. Ideally, if the system is in
a legitimate state, only one process holds a token in the configuration.
However, errors may occur when the system enters into a multiple token
state. Herman's algorithm is a randomized protocol to reach a one-token
state after an error, hence the name self-stabilizing; cf. \cite{D74,D00}.

The method of the algorithm is the following: in every step of the discretely treated time, if a process holds a token then it keeps it with probability $\frac{1}{2}$ or passes it to its clockwise neighbor with probability $\frac{1}{2}$, independently of the other token-passes. 
If a process kept its token in a step but also receives one, then both tokens disappear. 
By the implementation of the processes, we can guarantee that Herman's algorithm starts at a configuration where there is an odd number of tokens, hence the mentioned algorithm will eventually yield a one-token state with probability 1. 
We note that it is just as reasonable from a mathematical point of view to start with an even number of tokens and run the process until all tokens disappear. 
Interestingly, this setup is much easier to handle, and was solved in \cite{FZ15}. 

Several questions arise naturally about the distribution of the execution
time of self-stabilization, i.e., the hitting time $\mathbf{T}$ of a one-state
configuration \cite{KMOWW12}. Since the complete description of the
distribution $\mathbb{P}(\mathbf{T}\geq t)$ did not turn out to be an
accessible problem, the analysis focused mainly on the derived quantity
$\mathbb{E}(\mathbf{T})$. The denominator, Herman, proved the upper bound $\frac{1}{2}N^{2}\log N$
on $\mathbb{E}(\mathbf{T})$ in the original paper \cite{H90}, which was improved
to $O(N^{2})$ by multiple authors independently \cite{FZ15,FMP05,MM05,N05}.

To find a tight bound, it is reasonable to search for the extremum
of $\mathbb{E}(\mathbf{T})$ as a function of the initial configuration of
the tokens. Assuming that the stabilization process starts with three tokens,
the maximum of $\mathbb{E}(\mathbf{T})$ is realized on the equidistant starting
position of the tokens (or the closest configuration to that, if $N$
is not divisible by 3). This is a consequence of the description of
$\mathbb{E}(\mathbf{T})$ given by \cite{MM05} for all the initial configurations
with three tokens. They found an explicit formula for $\mathbb{E}(\mathbf{T})$
in terms of the ``distances'' of the tokens, where by distance of
the tokens $X_{1}$ and $X_{2}$ we mean the length of the arc connecting
$X_{1}$ and $X_{2}$ avoiding the third token $X_{3}$. Given
these distances $a$, $b$, $c\in\mathbb{N}$ of the tokens (where necessarily
$a+b+c=N$ by definition), the expectation of $\mathbf{T}$ can be
expressed as
\[
\mathbb{E}(\mathbf{T})=\frac{4abc}{N}
\]
This expression clearly has the maximum at the states where $a,b,c$ are the
nearest integers around $\frac{N}{3}$ summing up to $N$. In particular,
$a=b=c=\frac{N}{3}$ if $N$ is divisible by $3$. In \cite{MM05},
it was also conjectured to be the only maximum of $\mathbb{E}(\mathbf{T})$
considering all possible initial configurations, not necessarily
with three tokens. 

We give a proof to this conjecture by using a method that can be generalized in many directions. 
In the original phrasing of the conjecture, in every round of the discrete process, each token either keeps its current position or makes a move in the positive direction, and both events occur with equal probability $p=\frac{1}{2}$. 
We treat the unbiased version where all tokens make a move independently with the same probability $p$; cf. \cite{F84,RS11}. 
The argument is not purely combinatorial in its intrinsic nature. 
Hence, it can be generalized to the case when the movement of each token is described by a Poisson process or when the circular graph is replaced by a continuous circle on which the tokens move by independent Brownian motions with the same parameters.

Moreover, we show that in the (unbiased) discrete version of the protocol with parameters $p,N$, and for $\varepsilon=4p(1-p)\sin^{2}\big(\frac{\pi}{2N}\big)$, we have 
$\mathbb{E}\bigg(\Big(\frac{1}{1-\varepsilon}\Big)^{\mathbf{T}}\bigg)\leq\frac{3}{2}$, with equality if and only if we start from the three-token equidistant configuration. 
Furthermore, subject to the the correctness of an optimization result that we tested by computer, the maximum of $\mathbb{E}\big(\alpha^{\mathbf{T}}\big)$ is attained at the three-token equidistant configuration for all $1\leq \alpha\leq \frac{1}{1-\varepsilon}$.
All the evidence point towards the potential result that $\mathbb{P}(\mathbf{T}\geq t)$ is maximized by the equidistant three-token configuration for all $t$. 
Indeed, $\mathbb{E}\big(\alpha^{\mathbf{T}}\big)$ is also a linear combination of the $\mathbb{P}(\mathbf{T}\geq t)$ just like $\mathbb{E}(\mathbf{T})$, but with
weights $\alpha^{t}-\alpha^{t-1}$
rather than $1$'s, as in the case of $\mathbb{E}(\mathbf{T})$. 
In \cite{KMOWW12}, a formula was established to $\mathbb{P}(\mathbf{T}\geq t)$
assuming that there are three tokens. 
As a consequence, they have shown that the maximum of $\mathbb{P}(\mathbf{T}\geq t)$ is indeed attained at the equidistant three-token starting state when we consider the three-token initial states. The next step could be to obtain this theorem with no restriction on the number of tokens.

%The authors are thankful to Andrzej Murawski who drew the attention to the Herman Protocol Conjecture at the first problem solving session of the DIMAP Retreat, in March 2013. 
%The key idea of this solution, Theorem \ref{thm:weaker theorem} and the sketch of the modification leading to Theorem \ref{thm:main theorem} was explained in the second Problem solving session. 
%Subsequently, another solution of the conjecture was published in \cite{BGKOW15} using a different method.
This paper is an extended and improved version of the manuscript \cite{CsM}, which proved the Herman Protocol Conjecture parallelly with and independently of \cite{BGKOW15}.

\section{Main results}
\subsection{Basic setup and three-token states}
We implement a somewhat modified viewpoint on the described processes. 
Since the arguments require some symmetry, for our purposes it is better to rotate the base space by $\frac{2\pi}{2N}$ after every step counter-clockwise, where $N$ stands for the number of nodes. 
This slight notational modification have the effect that the number of nodes gets doubled, but half of them is necessarily avoided by the tokens in every step. 
At least this is the case in the classical, discrete version of the protocol, but not necessarily in the other variants we consider in this paper. 
In the sequel we refer to $2N$ as the number of nodes. 
Moreover, the tokens now move in a symmetrized way. 
In the standard Herman protocol, they either move to the clockwise neighboring new node with probability $\frac{1}{2}$ or move in the opposite direction to the counter-clockwise neighbor with probability $\frac{1}{2}$, all independently. 

We now generalize this setup by choosing another parameter (besides $N$). 

\begin{defn}
Given a $p\in ]0,1[$, the discrete Herman protocol with parameters $p,N$ is the process where a number of tokens are moved independently along a circle with $2N$ nodes, such that each token moves to its counter-clockwise neighboring node with probability $p$, and to the other neighbor with probability $1-p$, all independently. 
In other words, tokens are taking independent biased random walks $\pmod {2N}$. 
\end{defn}

Using this reformulation, it is natural to consider further variants. 
We define two more setups that we can handle in essentially the same way as the classical discrete version. 
In these new variants, the movement of each token is a continuous time process, in fact, a L\'evy process. 
It is vital in some arguments that the tokens cannot jump over one another. 
Hence, if we intend to preserve the symmetry between the tokens in the sense that the processes describing their movements are independent copies of the same L\'evy process, only the following two special types can be considered. 

\begin{defn}
In the exponential clock (or Poisson) variant, the tokens are still positioned on the $2N$ nodes along the circle, and steps are discrete. Each token moves to its neighbor in the positive direction with probability $p$, and to the other one with probability $1-p$. 
However, the timeline is continuous, and each token has a corresponding exponential clock with mean $1$: whenever it goes off, the token takes a step. 
\end{defn}

\begin{defn}
The Brownian (or Wiener) variant is continuous, making the nodes irrelevant. 
The tokens can be positioned anywhere around the circle with perimeter $2N$, and they each independently move via a Brownian motion with variation $1$. 
In this setup, $N$ need not be an integer. 
\end{defn}

If one were only interested in the Brownian variant, a somewhat more natural parametrization could be chosen. However, in order to be consistent with the other variants, and to easily compare the three situations, we use this one. 

\begin{figure}[H]
\textit{\quad\includegraphics[scale=0.5]{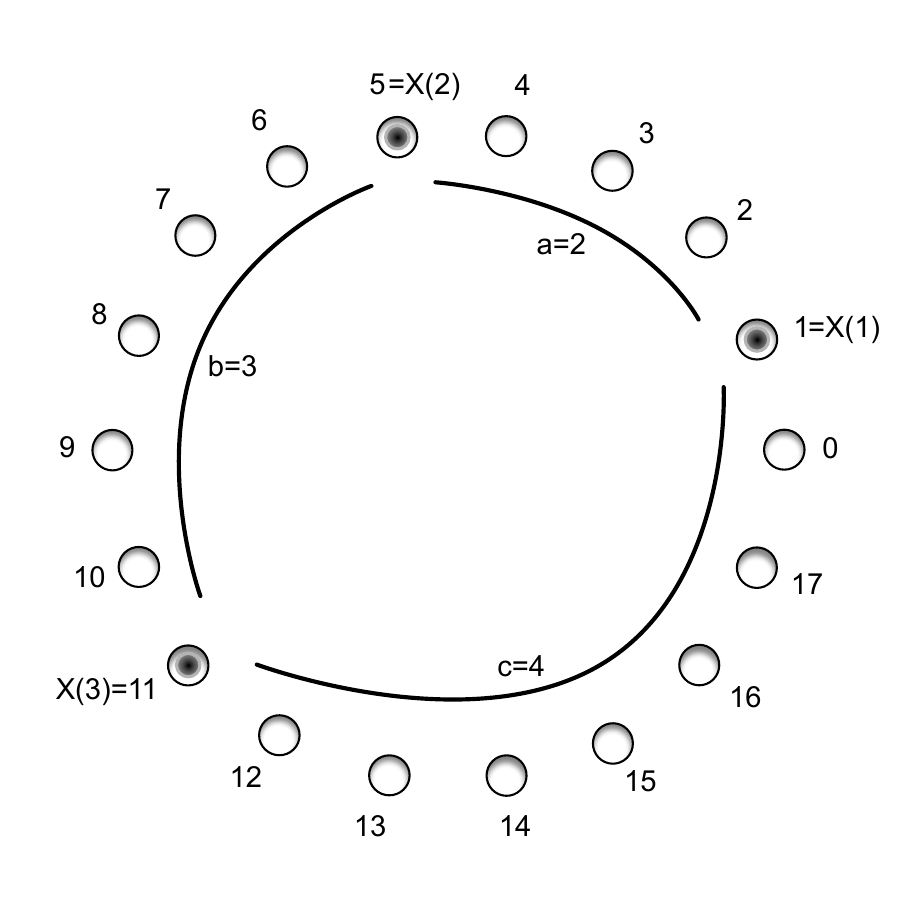}\quad\quad\quad\quad\includegraphics[scale=0.5]{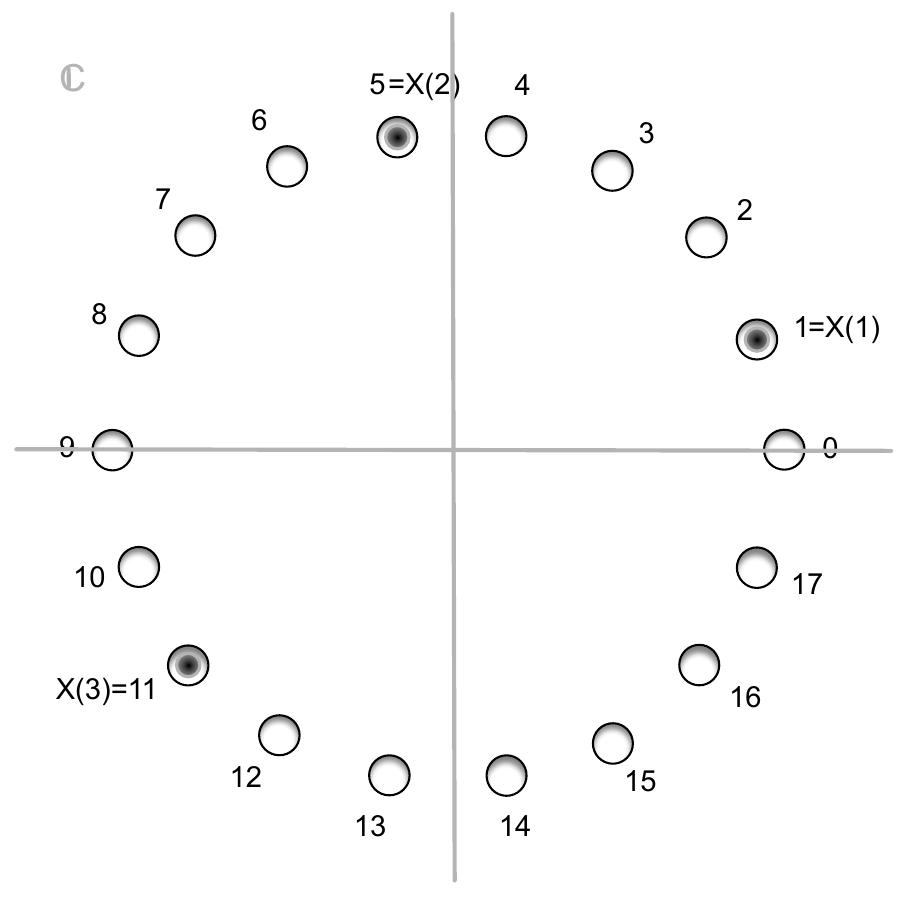}}
\protect\caption{\textit{Illustration of a three-token configuration with $N=9$. On the left, the distances between the tokens are indicated. On the right, the same configuration is placed in the complex plane.}}
\end{figure}\label{fig:distandcomp}

The three-token states play a crucial role in the argument, thus we first determine the expected time to absorption from  such initial states. 
The formula in the unbiased discrete setup ($p=\frac{1}{2}$) was shown in \cite{MO05}. 
In the standard discrete protocol, a three-token state was usually represented by three numbers $a$, $b$, $c$, the distances between the three pairs of tokens, i.e. the number of nodes on the arcs connecting two tokens while avoiding the third. 
Then the expected runtime of the process is $\frac{4abc}{N}$ according to \cite{MO05}.
Note that in our modified setup with the circle having perimeter $2N$ rather than $N$, we require a slight adjustment to get the same formula: the distances $a$, $b$, $c$ are now half the arc lengths connecting each pair of tokens, as the nodes are doubled. 
So in order to obtain $\frac{abc}{27}$ for the expected runtime for three-token states, in our formulation with the $2N$ nodes and each arc between neighboring nodes having length 1, the actual arc lengths between the tokens are $2a$, $2b$, $2c$, respectively. 
In order to keep the consistency, we still refer to the distances as $a$, $b$, $c$; see the left hand side of Figure~\ref{fig:distandcomp}. 
This is a common feature for all three variants considered in the paper. 
Throughout the paper, we make several references to the \emph{equidistant three-token configuration} without elaborating on the cases where $2N$ is not divisible by $3$. 

\begin{lem}\label{lem:time_absorption}
Let $x$ be an initial three-token state with distances $a$, $b$, $c$ between the three pairs of tokens. Then the Herman protocol is expected to terminate in $\mathbb{E}(\mathbf{T}(x))=\mu\frac{abc}{N}$ steps, where 
\begin{itemize}
\item $\mu=\frac{1}{p(1-p)}$ for the discrete version with parameters $p$, $N$;
\item $\mu=4$ for the exponential clock variant with parameters $p$, $N$;
\item $\mu=4$ for the Brownian process with parameter $N$.
\end{itemize}

In particular, we have the upper bound $\mathbb{E}(\mathbf{T}(x))\leq \frac{\mu N^2}{27}$ for all three-token states $x$, with equality if and only $x$ is the equidistant three-token configuration. 
Moreover, $\mathbb{E}(\mathbf{T})$ is finite for all three variants and arbitrary initial states. 
\end{lem}
\begin{proof}
The system of linear equations obtained from first step analysis has a unique solution. 
Hence, to solve the discrete case, it is enough to check that the formula $\frac{abc}{p(1-p)N}$ is consistent with the Law of Total Expectation. 
That is, the value of the expression $\frac{abc}{p(1-p)N}$ drops by 1 in average after one step. 
After the first step, the process can be in eight different states. 
E.g., if the tokens $A$, $B$ move counter-clockwise and $C$ moves clockwise, then $2a'=2a-2$, $2b'=2b+2$, $2c'=2c$, making the new distances $a'=a-1$, $b'=b+1$, $c'=c$ after one step. 
The weighted average of values of the above formula is 
\[
p^3\frac{abc}{p(1-p)N} + p^2(1-p)\frac{a(b-1)(c+1)}{p(1-p)N} + p^2(1-p)\frac{(a+1)b(c-1)}{p(1-p)N} + p^2(1-p)\frac{(a-1)(b+1)c}{p(1-p)N}+
\]
\[
+p(1-p)^2\frac{a(b+1)(c-1)}{p(1-p)N} + p(1-p)^2\frac{(a-1)b(c+1)}{p(1-p)N} + p(1-p)^2\frac{(a+1)(b-1)c}{p(1-p)N} + (1-p)^3\frac{abc}{p(1-p)N} =
\]
\[
= \frac{abc}{p(1-p)N} - (p^2(1-p)+p(1-p)^2)\frac{a+b+c}{p(1-p)N} = \frac{abc}{p(1-p)N} - \frac{a+b+c}{N} = \frac{abc}{p(1-p)N} -1
\]

The continuous-time variants can be dealt with similarly. 
If we have an exponential clock, then under a small amount of time $\Delta t$, the probability that a clock goes off at least twice or at least two clocks go off is $O(\Delta t^2)$. 
The probability for a clock to ring (at least or exactly) once is $\Delta t e^{-\Delta t}+O(\Delta t^2)= \Delta t +O(\Delta t^2)$. 
Thus there are essentially seven events to consider that have more than $O(\Delta t^2)$ probability: no clock goes off, or exactly one out of the three clocks rings and the token moves into one out of two possible neighbors. 
When only one token moves, say $A$ in the clockwise direction, then the new distances are $a'=a$, $b'=b-\frac{1}{2}$, $c'=c+\frac{1}{2}$. 
This leads to the following estimate of the expression $\frac{4abc}{N}$ after one step. 

\[
(1-3\Delta t)\frac{4abc}{N} + (1-p)\Delta t\frac{4a(b-\frac{1}{2})(c+\frac{1}{2})}{N} + (1-p)\Delta t\frac{4(a+\frac{1}{2})b(c-\frac{1}{2})}{N} + (1-p)\Delta t\frac{4(a-\frac{1}{2})(b+\frac{1}{2})c}{N}+
\]
\[
+p\Delta t\frac{4a(b+\frac{1}{2})(c-\frac{1}{2})}{N} + p\Delta t\frac{4(a-\frac{1}{2})b(c+\frac{1}{2})}{N} + p\Delta t\frac{4(a+\frac{1}{2})(b-\frac{1}{2})c}{N} +O(\Delta t^2)= 
\]
\[
= \frac{4abc}{N} - \Delta t \frac{a+b+c}{N} +O(\Delta t^2) = \frac{4abc}{N} - \Delta t +O(\Delta t^2)
\]

Thus as $\Delta t \rightarrow 0$, we have that the derivative of $\frac{4abc}{N}$ is $-1$ in expectation. 

Similarly, if the motion of tokens are independent Wiener processes, then let $A$, $B$, $C$ denote the original position of the tokens. 
After a small amount of time $\Delta t$, the new position of the tokens are $A+W_A$, $B+W_B$, $C+W_C$, where $W_A$, $W_B$, $W_C\sim {\text{Gaussian}}(0, \Delta t)$, with a negligible error due to the circular nature of the processes. 
Hence, the new distances are $a'=a+\frac{1}{2}(W_C-W_B)$, $b'=b+\frac{1}{2}(W_A-W_C)$, $c'=c+\frac{1}{2}(W_B-W_A)$ with a negligible error, making the expression $\frac{4abc}{N}$ after a small time $\Delta t$ equal to

\[
\frac{4}{N}\left(a+\frac{1}{2}(W_C-W_B)\right)\left(b+\frac{1}{2}(W_A-W_C)\right)\left(c+\frac{1}{2}(W_B-W_A)\right) = 
\]
\[
= \frac{4abc}{N} + \frac{1}{N}\left(a(W_A-W_C)(W_B-W_A) + b(W_B-W_A)(W_C-W_B) + c(W_C-W_B)(W_A-W_C) \right) 
\]
which after taking expectation, and keeping in mind that $W_A$, $W_B$, $W_C$ are i.i.d. Gaussian variables with mean zero and variance $\Delta t$ is:
\[
\mathbb{E}\left(\frac{4abc}{N} + \frac{1}{N}\left(a(W_A-W_C)(W_B-W_A) + b(W_B-W_A)(W_C-W_B) + c(W_C-W_B)(W_A-W_C) \right)\right) = 
\]
\[
= \frac{4abc}{N} - \frac{1}{N}\left(a\mathbb{E}(W_A^2) + b\mathbb{E}(W_B^2) + c\mathbb{E}(W_C^2) \right) = \frac{4abc}{N} - \frac{a+b+c}{N}\Delta t = \frac{4abc}{N} - \Delta t 
\]
Thus as $\Delta t \rightarrow 0$, we have that the derivative of $\frac{4abc}{N}$ is $-1$ in expectation. 

The second assertion about initial configurations with three tokens is a simple application of the AM-GM inequality. 

Finally, given any initial configuration with more than one tokens, pick any three tokens. 
Then the expected time of first collision is at most the expected runtime of the process with initial configuration containing those three tokens. 
Thus as long as there are more than one tokens, in a time that is finite in expectation, there is a collision. 
The sum of $\frac{K-1}{2}$ variables with finite expected value also has a finite expected value. 
\end{proof}

The main goal of the paper is to generalize the upper bound $\mathbb{E}(\mathbf{T}(x))\leq \mu\frac{N^2}{27}$ to all possible initial states $x$ for all three variants of the protocol. 

\begin{thm}\label{thm:main theorem}
Let $\mathbf{T}(x)$ denote the hitting time of the one-token state starting from the initial state $x$ for any of the three variants of the Herman protocol. 
Then $\mathbb{E}(\mathbf{T}(x))\leq \frac{\mu N^2}{27}$, with equality if and only if $x$ is the equidistant three-token configuration, where $\mu=\frac{1}{p(1-p)}$ for the discrete variant, and $\mu=4$ for the exponential clock version and for the Brownian process. 
\end{thm}

To verify the bound in Theorem~\ref{thm:main theorem}, it seems natural not only to keep count on when the process is terminated, but to have a way to measure ``how far'' we are from the end in expectation. 
Then the goal becomes to show that this measure is the worst (i.e. the highest) throughout the whole process if and only if the initial state is the equidistant three-token configuration. 
To this end, we define two ``potentials'' that are expected to grow, and that start off and end up in $[0,1]$. 
The first such potential is $\Phi$, see Definition~\ref{def:Phipot}. 
It is only relevant for three-token states; it assigns to a three-token state $x$ the expected value of the remaining time until the process terminates starting from $x$, rescaled into $[0,1]$. 
Note that Lemma~\ref{lem:time_absorption} guarantees that $\Phi$ is indeed between 0 and 1. 

\begin{defn}\label{def:Phipot}
Let $x$ be an initial three-token state with distances $a$, $b$, $c$ between the three pairs of tokens. 
Then for all three variants of the Herman process we define $\Phi(x) = 1 - \mathbb{E}(\mathbf{T}(x))/\left(\mu\frac{N^2}{27}\right) = 1-\frac{27abc}{N^3}$. 
\end{defn}

\subsection{The other potential}

The following arguments are explained for the discrete variant of the process, but it is easy to see that everything generalizes to the other versions, as well. 
The only relevant nontrivial property that we make use of is that the tokens have a circular order that cannot change without a collision of tokens. 
That is, tokens cannot jump through one another without first occupying the same position; this was exactly the property that we focused on when defining the three variants. 

To define the non-trivial potential, the core idea of the paper, we number the nodes by $0$, $1$, $2$, $\dots$, $2N-1$. 
The location of the $j$-th token at time $t\in\mathbb{N}_{\geq0}$ is described by the random variable $X_{t}(j)$ where $j=1$, $2$, $\dots$, $K_{t}$ and $K_{t}$ stands for the number of tokens at time $t$, where the tokens are numbered compatible to their ordering on the circle (but the beginning of the enumeration is arbitrary). 
Generalizing the notation $K_{t}$ we will write $K_{t}(x)$ for the (random) number of tokens at time $t$ for the process starting at the initial state $x$. 
In particular, $K(x):=K_{0}(x)$ denotes the number of tokens at state $x$. 
As before, $\mathbf{T}:=\mathbf{T}(x):=\min\{t\ |\ K_{t}(x)=1\}$ is the hitting time of a one-token state, i.e. the execution time of the self-stabilizing algorithm. 
Note that this notion is not affected by the symmetrization of the process we implemented in the previous chapter. 
Also, we need a notation for the hitting time of a three-token state, as it turns out to be a crucial point in the evolution of the process, so we put $\tau:=\min\{t\ |\ K_{t}(x)=3\}$.

Just like $\Phi(x)$, the new potential $x\mapsto\Psi(x)\in[0,1]$ also measures how far our state is from the final state in expectation. 
The growth speed of $\Psi$ can be estimated without trying to compute the first potential $\Phi$ for all configurations with an arbitrary number of tokens, a seemingly impossible challenge. 
However, we can show that the two potentials are reasonably close to each other on three-token states; see Lemma~\ref{lem:interchange of potentials}. 

In the definition of $\Psi$, we use the complex exponential function $k\mapsto e^{\frac{2\pi i}{2N}k}=e^{\frac{\pi i}{N}k}$. 
This notation also implicitly contains an identification of the circle with the complex unit circle (the identification
was essentially chosen when we numbered the nodes); see the right side of Figure~\ref{fig:distandcomp}. 
This arbitrary choice could in principle cause some trouble. But as we will soon see, the potential $\Psi$ is invariant under rotation of the circle (i.e., the choice of the node with number 0), solving the issue; see Proposition~\ref{prop:properties of psi w proof}. 

\begin{defn}
Let $x$ be an arbitrary state and assume that $0\leq x(1)<x(2)<\dots<x(K)\leq 2N-1$
where $x(j)$ is the position of the $j$'th token of the state $x$ using counter-clockwise enumeration of the tokens starting at the direction $1\in\mathbb{C}$ (the node with number 0); see the right side of Figure~\ref{fig:distandcomp}. 
Then the potential $\Psi$ is 

\[
\Psi(x):=\Big|\sum_{j=1}^{K(x)}e^{\frac{\pi i}{N}\frac{1}{2}x(j)}(-1)^{j}\Big|^{2}
\]
\end{defn}

Geometrically, $x\mapsto\Psi(x)$ can be described as summing up the (directed) angle bisectors of the vectors $e^{\frac{\pi i}{N}x(j)}$ and the fixed unit vector $1$ with an extra twist. 
Namely, for odd $j$ we reflect the resulting angle bisector vector to the origin. 
Informally, this reflection is applied to stabilize the quantity under the disappearance of two colliding tokens. 
Formally, it means that if $x(j)=x(j+1)$ then deleting these two tokens from the vector $x$ does not change the value of $\Psi(x)$. 
Also the alternating sign is responsible for the independence of $\Psi$ from the choice of the direction, i.e., the identification of the circle with the unit circle in the complex plane. 

\begin{prop}\label{prop:properties of psi w proof} 
For any state $x$ we have \label{prop:properties of psi w/o proof}
\begin{enumerate}
\item The choice made at the identification of the plane with $\mathbb{C}$
does not affect $\Psi(x)$. That is, $\Psi(x)$ is invariant
under the simultaneous translation of the $x(j)$, even if during the
translation, a token jumps over $1\in\mathbb{C}$.
\item The disappearance of two colliding tokens does not affect $\Psi(x)$. 
\item $\Psi(x)\leq1$, with equality if and only if there is only one token
at state $x$.
\item $\Psi(x)\geq0$, with equality if and only if $x$ is an equidistant configuration with at least three tokens.
\end{enumerate}
\end{prop}
\begin{proof}
For the first statement, note that a $-2\pi\frac{k}{2N}$ rotation of the choice of $1\in\mathbb{C}$ corresponds to a multiplication of every vector by a length 1 complex number. 
Assuming that $0\leq x(j)$ implies $0\leq x(j)+k$ for all $j$, i.e., no token jumps over the abstract border between $0$ and $2N-1$, this transformation does not alter $\Psi(x)$ since
\[
\Big|\sum_{j=1}^{K(x)}e^{\frac{\pi i}{N}\frac{1}{2}x(j)}(-1)^{j}\Big|^{2}=\Big|\sum_{j=1}^{K(x)}e^{\frac{\pi i}{N}\frac{1}{2}(x^{*}(j)+k)}(-1)^{j}\Big|^{2}=\Big|e^{\frac{\pi i}{N}\frac{1}{2}k}\cdot\sum_{j=1}^{K(x)}e^{\frac{\pi i}{N}\frac{1}{2}x^{*}(j)}(-1)^{j}\Big|^{2}=\Big|\sum_{j=1}^{K(x)}e^{\frac{\pi i}{N}\frac{1}{2}x^{*}(j)}(-1)^{j}\Big|^{2}
\]
where $x^{*}$ stands for the renumbered sequence. 
So it is enough to consider the case when there is a (special) token at the node with index 0, i.e., $x(1)=0$, and we rotate the nodes and tokens by the angle $-2\pi\frac{1}{2N}$. 
That rotation changes several things. 
The enumeration of tokens is altered, leading to the change in the sign $(-1)^j$ for all the vectors in the sum except for the special one corresponding to the token that ends up in $2N-1$: both its original and new sign $(-1)^j$ in the sum is negative, as $(-1)^1=(-1)^{K}=-1$. 
Moreover, the value $x(j)$ is decreased by 1 for all $j$ except for the one corresponding to the special token, which increases from 0 to $2N-1$. 
Therefore, every summand in the expression $\sum\limits_{j=1}^{K(x)}e^{\frac{\pi i}{N}\frac{1}{2}x(j)}(-1)^{j}$ which does not correspond to the special token is multiplied by $-e^{-\frac{\pi i}{2N}}$. 
The special summand is multiplied by $e^{\frac{\pi i}{2N}\cdot (2N-1)}=-e^{-\frac{\pi i}{2N}}$, the same factor. 
As this complex number has modulus 1, the value $\Big|\sum_{j=1}^{K(x)}e^{\frac{\pi i}{N}\frac{1}{2}x(j)}(-1)^{j}\Big|^{2}$ is unaltered, verifying the first item of the assertion.

The second statement is clear, since if $x(j)=x(j+1)$ then they represent the same vector contributing to the sum with different signs, so their sum is zero:
\[
e^{\frac{\pi i}{N}\frac{1}{2}x(j)}(-1)^{j}+e^{\frac{\pi i}{N}\frac{1}{2}x(j+1)}(-1)^{j+1}=0
\]
and the change of the enumeration (indices after $j+1$ decrease by $2$) does not affect the sum as $(-1)^{\ell}=(-1)^{\ell-2}$.

To prove the third statement, we may assume by the first item that $x\big(1\big)=0$. 
Then we can partition the remaining tokens into $\frac{K(x)-1}{2}$ consecutive pairs so that the sum of the corresponding vectors can be expressed as
\[
\Big|\sum_{j=1}^{K(x)}e^{\frac{\pi i}{N}\frac{1}{2}x(j)}(-1)^{j}\Big|^{2}=\Big|1+\sum_{j=2}^{K(x)}e^{\frac{\pi i}{N}\frac{1}{2}x(j)}(-1)^{j}\Big|^{2}=\Big|1+\sum_{k=1}^{\frac{K(x)-1}{2}}\big(e^{\frac{\pi i}{N}\frac{1}{2}x(2k)}-e^{\frac{\pi i}{N}\frac{1}{2}x(2k+1)}\big)\Big|^{2}
\]
where each $e^{\frac{\pi i}{N}\frac{1}{2}x(2k)}-e^{\frac{\pi i}{N}\frac{1}{2}x(2k+1)}$
can be pictured as the vector starting at the end-point of
the angle bisector unit vector $e^{\frac{\pi i}{N}\frac{1}{2}x(2k+1)}$
and ending at the endpoint of $e^{\frac{\pi i}{N}\frac{1}{2}x(2k)}$,
so it is a chord of the unit-circle. 
It is clear that the sum of several directed chords on a half-circle plus the vector $(1,0)$ will not sum up to vector longer than $1$. 
Indeed, the sum of their projection to any line in the plane has length at most 1, which is equivalent to having length at most 1. 
The case of equality means that there is a projection such that this sum of vectors has length 1 in one direction. 
But as the chords must not meet (since $x(j)\neq x(j+1)$ for all $j$) and also no angle bisector can end at $-1$, it is possible only if that projection is on a horizontal line, so there is no token besides the one at the node $0$.

The inequality in the fourth item is obvious, since the modulus of any complex number is non-negative. 
The case of equality can be investigated by the same geometric argument as in the third item. 
Namely, a sum of vectors is the zero vector if and only if all of its projections are zero. 
\end{proof}

Now, we fix the initial state $x$ of the process $t\mapsto X_{t}$.
Let us denote by $Y_{t}=\Psi(X_{t})$ the value of the potential defined above on the random process at time $t$. 
As usual,  $\mathcal{F}_{t}:=\sigma(X_{t}(j)\ |\ s\leq t,\ j\leq K_{0})$ is the standard filtration of the process, and the number of tokens $K_{t}=K_{t}(x)$ at time $t$ was defined earlier. 

The evolution of $Y_{t}$ is described by the following lemma.

\begin{lem}\label{lem:recursion w proof} \quad
\begin{itemize}
\item Given the discrete version of the Herman protocol with parameters $p, N$, let $\varepsilon = 4p(1-p)\sin^2\left(\frac{\pi}{2N}\right)$. Then for any $t\in\mathbb{N}$ we have 
\[
\mathbb{E}(Y_{t+1}-Y_t\mid\mathcal{F}_{t})=\varepsilon(K_{t}-Y_{t})
\]
\item In the exponential clock variant, let $\varepsilon = 4\sin^2\left(\frac{\pi}{4N}\right)$. Then $\lim\limits_{\Delta t \rightarrow 0} \frac{1}{\Delta t}\mathbb{E}(Y_{t+\Delta t}-Y_t\mid\mathcal{F}_{t})=\varepsilon(K_{t}-Y_{t})$. 
\item In the Brownian version, let $\varepsilon = \frac{\pi^2}{4N^2}$. Then $\lim\limits_{\Delta t \rightarrow 0} \frac{1}{\Delta t}\mathbb{E}(Y_{t+\Delta t}-Y_t\mid\mathcal{F}_{t})=\varepsilon(K_{t}-Y_{t})$. 
\end{itemize}
\end{lem}
\begin{proof}
Let $\delta_t(j, \Delta t) = X_{t+\Delta t}(j) - X_t(j)$ be the increment of the position of the $j$-th token at time $t$ measured in the signed distance on the circle. In particular, in the discrete variant, $\delta_t(j, 1) = X_{t+1}(j) - X_t(j)$ is $\pm 1$: it is $+1$ if the $j$-th token moved in the positive direction, and $-1$ if the $j$-th token moved in the negative direction. 
The choice $\Delta t=1$ in the discrete case is recommended as in one move, tokens cannot jump over one another, hence the ordering of tokens after this particular amount of time is the same along the circe. 
In the continuous variants, such small amount of time is not possible to choose: in principle, any token could jump over another in any positive amount of time if colliding tokens were not removed. 
However, the error introduced by this imprecision is negligible if $\Delta t$ is sufficiently small; much smaller than the distance of the two closest tokens at time $t$. 
Note that the distribution of $\delta_t(j, \Delta t)$ only depends on the variant of the process considered and the choice of $\Delta t$: at any given time $t$ for any given index $j$, the increment of the $j$-th token has a fixed distribution for all three processes, provided that we disregard the negligibly small possibility of collision with another token before the $\Delta t$ amount of time, and allow the token to keep moving if that happens. So we should let the tokens make their move, and take stock after the $\Delta t$ amount of time, removing those that should have been removed. 
Let $\delta$ be the distribution of the difference of two independent samples from $\delta_t(j, \Delta t)$: in order to simplify the notation, we do not indicate its dependence on $\Delta t$. 
As $\Delta t$ can be considered fixed for all our purposes, this should not cause any confusion in the sequel. 
In other words, $\delta$ is the distribution $\delta_t(a, \Delta t)- \delta_t(b, \Delta t)$ for two different indices $a,b$ at any given time $t$. 

In the discrete version, $\delta$ is the difference of two Rademacher distributions with parameter $p$ (here, $\Delta t = 1$). 
That is, $\mathbb{P}(\delta=0) = 1-2p+2p^2$ and $\mathbb{P}(\delta=1) = \mathbb{P}(\delta=-1) = p-p^2$, with characteristic function $\chi_\delta(s) = 1-4p(1-p) \sin^2(s)$. 

In the exponential clock version, under an appropriately small time $\Delta t$, a token stays still with probability $1-\Delta t + O(\Delta t^2)$, makes one move in the positive direction with probability $p\Delta t + O(\Delta t^2)$, and makes one move in the negative direction with probability $(1-p)\Delta t + O(\Delta t^2)$; every other possible move has negligible probability. 
Hence, $\mathbb{P}(\delta=0) = 1-2\Delta t + O(\Delta t^2)$, $\mathbb{P}(\delta=1) = \mathbb{P}(\delta=-1) = \Delta t + O(\Delta t^2)$, and the other possibilities can be ignored. 
The characteristic function of the difference is $\chi_\delta(s) = 1- \Delta t \cdot 4\sin^2(s/2) + O(\Delta t^2)$. 

Finally, in the Brownian variant, the increments have normal distribution with mean 0 and variance $\Delta t$, thus the difference of two such distributions is once again normal with mean 0 and variance $2\Delta t$. 
Hence, $\chi_\delta(s) = e^{-\Delta t \cdot s^2} = 1-\Delta t \cdot s^2 + O(\Delta t^2)$. 

Now we are ready to compute $\mathbb{E}(Y_{t+1}-Y_t\mid\mathcal{F}_{t})$ with essentially the same calculation for all three versions of the protocol. 

\[
\mathbb{E}(Y_{t+1}-Y_t\mid\mathcal{F}_{t})= \mathbb{E}(\Psi(X_{t+\Delta t}))\mid\mathcal{F}_{t}) -Y_t= 
\]
\[= \mathbb{E}\left(\left(\sum\limits_{a=1}^{K_t} (-1)^ae^{\frac{\pi i}{2N} X_{t+\Delta t}(a)}\right)\left(\sum\limits_{b=1}^{K_t} (-1)^be^{-\frac{\pi i}{2N}X_{t+\Delta t}(b)}\right)\mid\mathcal{F}_{t}\right) - Y_t =
\]
\[
= \mathbb{E}\left(\left(\sum\limits_{a=1}^{K_t} (-1)^ae^{\frac{\pi i}{2N} (X_t(a)+\delta_t(a, \Delta t))}\right)\left(\sum\limits_{b=1}^{K_t} (-1)^be^{-\frac{\pi i}{2N}(X_t(b)+\delta_t(b, \Delta t))}\right)\mid\mathcal{F}_{t}\right) - Y_t =
\]
\[
= \mathbb{E}\left(\sum\limits_{a,b=1}^{K_t} (-1)^{a+b}e^{\frac{\pi i}{2N} (X_t(a)-X_t(b)+\delta_t(a, \Delta t)-\delta_t(b, \Delta t))}\mid\mathcal{F}_{t}\right) - Y_t
\]

Note that we may assume that there are still $K_t$ tokens at time $t+\Delta t$. Indeed, in the discrete version, the expression is not altered if we forget to remove colliding tokens according to item 2. of Proposition~\ref{prop:properties of psi w proof}, and in the continuous time variants, the probability of collision is negligible if $\Delta t$ is small. 
The expression $\delta_t(a, \Delta t)-\delta_t(b, \Delta t)$ appeared in the last exponent. 
We need to make a case distinction here: if $a=b$, then it is 0, otherwise it is a random variable with the same distribution as $\delta$. In the latter case when $a\neq b$, the variable $\delta_t(a, \Delta t)-\delta_t(b, \Delta t)$ is independent from $X_t(a)$ and $X_t(b)$, thus we can refer to the multiplicativity of the expected value for independent random variables. 
We transform the final expression in the calculation as if all pairs $a,b$ were different, and then we compensate for those $K_t$ pairs where $a=b$. 

\[
\mathbb{E}(Y_{t+1}-Y_t\mid\mathcal{F}_{t})= \mathbb{E}\left(\sum\limits_{a,b=1}^{K_t} (-1)^{a+b}e^{\frac{\pi i}{2N} (X_t(a)-X_t(b)+\delta_t(a, \Delta t)-\delta_t(b, \Delta t))}\mid\mathcal{F}_{t}\right) - Y_t = 
\]
\[
\mathbb{E}\left(\sum\limits_{a,b=1}^{K_t} (-1)^{a+b}e^{\frac{\pi i}{2N} (X_t(a)-X_t(b))}\mid\mathcal{F}_{t}\right)\mathbb{E}\left(e^{\frac{\pi i}{2N} \delta}\right) + \left(1-\mathbb{E}\left(e^{\frac{\pi i}{2N} \delta}\right)\right)K_t - Y_t = 
\]
\[
= Y_t\chi_\delta\left(\frac{2\pi}{N}\right) + \left(1-\chi_\delta\left(\frac{2\pi}{N}\right)\right) K_t - Y_t = \left(1-\chi_\delta\left(\frac{2\pi}{N}\right)\right) (K_t-Y_t) = \varepsilon(K_t-Y_t)
\]
according to the formulas for characteristic functions computed above. 
\end{proof}

\begin{cor}\label{cor:recursion}
For all three versions of the Herman protocol we have $\mathbb{E}(Y_{\tau})\geq4\varepsilon \mathbb{E}(\tau)+Y_{0}$. 
\end{cor}
\begin{proof}
Observe that Lemma \ref{lem:recursion w proof}, item 3. of Proposition~\ref{prop:properties of psi w proof}, and the fact that in a state with more than three tokens there are at least five tokens imply
\[
\mathbb{E}(Y_{t+\Delta t} - Y_t|\mathcal{F}_{t})\geq \varepsilon \Delta t (5-1) = 4\varepsilon \Delta t
\]
whenever $t<\tau$. 
Here, $\Delta t=1$ for the discrete variant and an infinitesimally small amount of time in the other two versions. 
Hence, the process 

$$Z_t= \begin{cases}
Y_t - 4\varepsilon t \text{\,\, if\,\,} t<\tau\\
Y_\tau - 4\varepsilon \tau \text{\,\, if\,\,} t\geq \tau
\end{cases}
$$

is a submartingale. 
To show that $Z_t$ is indeed integrable, first note that $|Y_t|\leq 1$ yields the trivial bound $|Z_t|\leq 1+4\varepsilon t$. 
In particular, if $t/\mathbb{E}(\tau)\leq 1$ then $\mathbb{E}|Z_t|\leq 1 + 4\varepsilon\cdot \mathbb{E}(\tau)$; note that $\mathbb{E}(\tau)$ is finite according to Lemma~\ref{lem:time_absorption}. 
If $t/\mathbb{E}(\tau)=C>1$ then we can apply the Law of Total Expectation and the Markov inequality to obtain 
\[\mathbb{E}|Z_t| = \mathbb{P}(t<\tau)\cdot \mathbb{E}\left(|Z_t| \mid t<\tau\right) +\mathbb{P}(t\geq \tau)\cdot \mathbb{E}\left(|Z_t| \mid t\geq\tau\right) \leq \frac{1}{C}(1+4\varepsilon C\cdot \mathbb{E}(\tau)) + 1\cdot \mathbb{E}|Y_\tau-4\varepsilon \tau| \leq
\]
\[
\leq   \frac{1}{C}(1+4\varepsilon C\cdot \mathbb{E}(\tau)) + \mathbb{E}(1+4\varepsilon \tau) \leq \frac{1}{C} + 4\varepsilon \cdot \mathbb{E}(\tau) + 1 + 4\varepsilon\cdot \mathbb{E}(\tau) \leq 2+8\varepsilon\cdot \mathbb{E}(\tau)
\]
Hence, according to the Optional Stopping Theorem $\mathbb{E}(Z\tau) \geq \mathbb{E}(Z_0)$, and consequently, $\mathbb{E}(Y_{\tau}) - 4\varepsilon \mathbb{E}(\tau)\geq Y_{0}$. 

\end{proof}

Informally, this corollary means that the potential $\Psi$ grows fast enough until we hit a three-token state. 
After time $\tau$ however, it slows down, as we can only guarantee a $2\varepsilon\cdot \Delta t$ growth under a (small) time period of length $\Delta t$ by the same argument as in the proof of Corollary~\ref{cor:recursion}. 
Fortunately, we have an exact formula to the other potential $\Phi$ for three-token states, see Definition~\ref{def:Phipot}. 
As we will see, it is easy to find an exact formula for $\Psi$ on three-token states, as well. 
So the vague idea is to estimate the growth of the potential $\Psi$ before the hitting time $\tau$ of three-token states by Corollary~\ref{cor:recursion}, and then switch to the other potential $\Phi$. 
In order to show that this switch can be carried out without a major loss in expectation, we need to compare the two potentials on three-token states. 

\begin{lem}\label{lem:interchange of potentials}
For any state $x$ with three tokens $\Phi(x)\geq 0.87\cdot\Psi(x)$.
\end{lem}
\begin{proof}
Let us denote the distance of the tokens of state $x$ (i.e. the number of original nodes on the arc connecting two tokens and avoiding the third) by $a$, $b$ and $c$ as in \cite{KMOWW12} and earlier on in the paper. 
So $a,b,c\in\{1,\dots,N\}$ (rather than $2N$) and $a+b+c=N$. Then $\Phi(x)=1-\frac{27abc}{N^{3}}$, and 
\[
\Psi(x)=\Big|(-1)e^{\frac{\pi i}{2N}x(1)}+e^{\frac{\pi i}{2N}x(2)}-e^{\frac{\pi i}{2N}x(3)}\Big|^{2}=
\]
\[
=3-2\mathrm{Re}\Big(e^{\frac{\pi i}{2N}\big(x(1)-x(2)\big)}\Big)-2\mathrm{Re}\Big(e^{\frac{\pi i}{2N}\big(x(2)-x(3)\big)}\Big)+2\mathrm{Re}\Big(e^{\frac{\pi i}{2N}\big(x(3)-x(1)\big)}\Big)=
\]
\[
=3-2\cos\Big(\frac{\pi}{2N}\big(x(1)-x(2)\big)\Big)-2\cos\Big(\frac{\pi}{2N}\big(x(2)-x(3)\big)\Big)+2\cos\Big(\frac{\pi}{2N}\big(x(3)-x(1)\big)\Big)=
\]
\[
=3-2\cos\Big(\frac{\pi a}{N}\Big)-2\cos\Big(\frac{\pi b}{N}\Big)+2\cos\Big(\frac{\pi c}{N}\Big)
\]
Notice that both expressions vanish at $a=b=c=\frac{N}{3}$ so we re-parametrize
them: let $u_1=\frac{a}{N}-\frac{1}{3}, u_2=\frac{b}{N}-\frac{1}{3}, u_3=\frac{c}{N}-\frac{1}{3}$. 
Then $u_i \in\Big[-\frac{1}{3},\frac{2}{3}\Big]$ and $u_1+u_2+u_3=0$. By the Newton-Girard formulas, the potential $\Phi$ is 
\[
\Phi(u_1,u_2,u_3)=1-27\Big(\frac{1}{3}+u_1\Big)\Big(\frac{1}{3}+u_2\Big)\Big(\frac{1}{3}+u_3\Big)= 1-(1-3u_1)(1-3u_2)(1-3u_3)=
\]
\[
=1-1+3(u_1+u_2+u_3)-9(u_1u_2+u_2u_3+u_3u_1)+27u_1u_2u_3=
\]
\[
=-\frac{9}{2}\left((u_1+u_2+u_3)^2-(u_1^2+u_2^2+u_3^2) \right) + 9\left(\frac{(u_1+u_2+u_3)^3}{2}-\frac{3(u_1+u_2+u_3)(u_1^2+u_2^2+u_3^2)}{2}+(u_1^3+u_2^3+u_3^3) \right)=
\]
\[
=\frac{9}{2}(u_1^2+u_2^2+u_3^2)+9(u_1^3+u_2^3+u_3^3)
\]
While the other potential is
\[
\Psi(u_1,u_2,u_3)=3-2\cos\left(\pi\left(u_1+\frac{1}{3}\right)\right)-2\cos\left(\pi\left(u_2+\frac{1}{3}\right)\right)-2\cos\left(\pi\left(u_3+\frac{1}{3}\right)\right)
\]
We divide by 9 for convenience, and show the estimate $\frac{1}{9}\Phi- \frac{0.87}{9} \Psi\geq 0$ by proving that the minimum of the function $\frac{1}{9}\Phi- \frac{0.87}{9}\Psi$ is attained at $(0,0,0)$. 
We apply the method of Lagrange multipliers to verify this claim. 
Note that the function is smooth and it is defined on a compact set, namely the triangle with vertices $(-\frac{1}{3},-\frac{1}{3},\frac{2}{3})$, $(-\frac{1}{3},\frac{2}{3},-\frac{1}{3})$, $(\frac{2}{3},-\frac{1}{3},-\frac{1}{3})$, hence the minimum exists. 
Moreover, on the border (the sides of the triangle), both $\Phi$ and $\Psi$ are constant 1, thus $\frac{1}{9}\Phi- \frac{0.87}{9}\Psi=\frac{0.13}{9}$. 
In contrast, the function attains the value 0 at the origin; thus the global minimum must be at an inner point of the domain.  

The Lagrange function for $\frac{1}{9}\Phi- \frac{0.87}{9} \Psi$ with the condition $u_1+u_2+u_3=0$ is 
$$L(u_1,u_2,u_3,\lambda)=\frac{1}{9}\Phi(u_1,u_2,u_3)-\frac{0.87}{9} \Psi(u_1,u_2,u_3)-\lambda(u_1+u_2+u_3)$$

The partial derivatives contain trigonometric functions, due to the cosine operations in the definition of $\Psi$. 
Leaving these in the formulas would present some technical difficulty. 
Hence, we re-parametrize the function for a final time: we prefer to work with the variables $x_i=-\cos\left(\pi\left(u_i+\frac{1}{3}\right)\right)$. 
On the one hand, this change of variables introduces a certain symmetry, as $x_i\in [-1,1]$. 
More importantly, the derivatives $\frac{d\, u_i}{d\, x_i}=\frac{d}{d\, x_i} \left(\frac{2}{3}-\frac{\arccos(x_i)}{\pi}\right) = \frac{1}{\pi\sqrt{1-x_i^2}}$, together with higher derivatives are rational functions of $x_i$ and $\sqrt{1-x_i^2}$, where the denominator is a power of the latter. 
As we are mainly interested in the sign of such an expression, and $\sqrt{1-x_i^2}$ is positive in the inner points of the domain, the computations are easier to carry out using this parametrization. 
Hence, 

$$L(x_1, x_2, x_3, \lambda) = \frac{1}{2}(u_1^2+u_2^2+u_3^2)-(u_1^3+u_2^3+u_3^3) - \frac{0.87}{9}(3+2x_1+2x_2+2x_3)-\lambda(u_1+u_2+u_3)$$ 

where $u_i$ is short for $\frac{2}{3}-\frac{\arccos(x_i)}{\pi}$. 

As usual, $\frac{\partial{L}}{\partial{\lambda}}=0$ yields the original condition $u_1+u_2+u_3=0$. 
The more interesting conditions gained from the other partial derivatives are $ \frac{1}{\pi\sqrt{1-x_i^2}}(u_i-3u_i^2-\lambda) - \frac{1.74}{9}=0$. 
By expressing $\lambda$ and putting $f(x)=u-3u^2-\frac{1.74\pi}{9}\sqrt{1-x^2}$, where once again $u$ is short for $\frac{2}{3}-\frac{\arccos(x)}{\pi}$, we obtain $f(x_1)=f(x_2)=f(x_3)$. 

We show that $f$ has a strictly monotone increasing segment on $]-1,0[$ followed by a decreasing one on $]0,1[$.  
The derivative of $f$ is $f'(x)=\frac{1}{\pi\sqrt{1-x^2}}(1-6u+\frac{1.74\pi^2}{9}x)$, thus $f'(0)=0$. 
In order to verify the claim, it is enough to show that $1-6u+\frac{1.74\pi^2}{9}x$ is positive on $]-1,0[$ and negative on $]0,1[$. 
Hence, it suffices to show that $1-6u+\frac{1.74\pi^2}{9}x$ is strictly decreasing. 
The derivative is 
\[
\left(1-6u+\frac{1.74\pi^2}{9}x\right)' = -\frac{6}{\pi\sqrt{1-x^2}}+\frac{1.74\pi^2}{9}\leq -\frac{6}{\pi}+\frac{1.74\pi^2}{9}\approx -0.0017<0
\] 
finishing the argument. 

In particular, every value is attained at most twice by $f$, and then by the pigeonhole principle, at least two of $x_1, x_2, x_3$ coincide. 
By the symmetry of the variables, we may assume that $x_1=x_2=x$ for some $x\in ]-1,1[$, making $u_1=u_2=u$ and $u_3=-2u$ for some $u\in ]-\frac{1}{3}, \frac{2}{3}[$. 

Hence, we can restrict our analysis to the section $\frac{1}{9}\Phi(u,u,-2u)-\frac{0.87}{9} \Psi(u,u,-2u)$, a univariate function with domain $[-\frac{1}{3}, \frac{1}{6}]$; indeed, since $-2u\geq -\frac{1}{3}$ we have $u\leq \frac{1}{6}$. 
By changing the variable as above, we have $x\in[-1,0]$ this time, as $u\leq \frac{1}{6}$ translates to $x\leq 0$. 
That is, the optimization problem is to find the minimum of $3u^2+6u^3-\frac{0.87}{9} \left(3+4x-2\cos\left(\pi\left(-2u+\frac{1}{3} \right) \right)\right)$. 
Note that $-2u+\frac{1}{3}= -2\left(u+\frac{1}{3}\right)+1$, leading to the simplification 
\[
-2\cos\left(\pi\left(-2u+\frac{1}{3} \right) \right)=2\cos\left(-2\pi\left(u+\frac{1}{3}\right)\right)=2\cos\left(2\pi\left(u+\frac{1}{3}\right)\right)=4\cos^2\left(\pi\left(u+\frac{1}{3} \right)\right)-2=4x^2-2
\] 
Consequently, the univariate function is $3u^2+6u^3-\frac{0.87}{9} \left(2x+1\right)^2$. 
After a division by 3, the assertion of the lemma reduces to the following claim: the function $g(x)=u^2+2u^3-\frac{0.29}{9} \left(2x+1\right)^2$ defined on $[-1,0]$ has a (unique) global minimum at $x=-\frac{1}{2}$, where $u=0$ and $g\left(-\frac{1}{2}\right)=0$. 
Note that $g\left(-1 \right)= g\left(0 \right)=\frac{0.13}{27}$ and $g\left(-\frac{1}{2} \right)=0$, showing that the minimum is at an inner point $x\in ]-1,0[$. 
The derivative of the function is 
\[
g'(x)=\frac{1}{\pi\sqrt{1-x^2}}(2u+6u^2)-\frac{1.16}{9}(2x+1)=
\frac{2}{\pi\sqrt{1-x^2}}\left(u+3u^2-\frac{0.58\pi\sqrt{1-x^2}}{9}(2x+1)\right)
\] 
Clearly $g'(-\frac{1}{2})=0$, and by L'H\^opital's rule $g'(-1)=\frac{1.16}{9}-\frac{2}{\pi^2}\approx -0.074<0$. 
In order to prove the claim, it is sufficient to verify that the sign of $g'$ is negative, 0, and positive on $[-1,-0.5[$, $\{-0.5\}$, and ]-0.5,0], respectively. 
For simplicity, we verify the analogous claim to the function $\frac{\pi\sqrt{1-x^2}}{2}g'(x)=u+3u^2-\frac{0.58\pi\sqrt{1-x^2}}{9}(2x+1)$, with the only difference that this function has value 0 at $-1$. 
Note how this problem is similar to the original claim regarding the function $g$, but we have reduced the degree of $u$. 
Obviously, if the revised claim holds, then the function must start off by a decreasing segment, followed by an increasing one, with the local minimum at $]-1,-0.5[$. 
It suffices to show that this is exactly the behavior of $\frac{\pi\sqrt{1-x^2}}{2}g'(x)$ on $]-1,0[$: that is, its derivative is negative on $]-1, \alpha[$ and positive on $]\alpha,0[$ for some $\alpha\in ]-1,-0.5[$. 

The derivative is 
\[
\left(\frac{\pi\sqrt{1-x^2}}{2}g'(x)\right)'=\frac{1}{\pi\sqrt{1-x^2}}\left(1+6u+\frac{0.58\pi^2}{9}x(2x+1) - \frac{1.16\pi^2}{9}(1-x^2)\right)=
\]
\[
=\frac{1}{\pi\sqrt{1-x^2}}\left(1+6u+\frac{0.58\pi^2}{9}(4x^2+x-2)\right)
\] 
This time, the limit at $-1$ is $-\infty$. 
Nevertheless, by omitting the factor $\frac{1}{\pi\sqrt{1-x^2}}$ again, we need to show that the resulting function $h(x)= 1+6u+\frac{0.58\pi^2}{9}(4x^2+x-2)$ has the behavior described above: negative on $]-1, \alpha[$ and positive on $]\alpha,0[$ for some $\alpha\in ]-1,-0.5[$. The limit $\lim\limits_{x\rightarrow (-1)+} h(x)$ is indeed negative, approximately $-0.364$. Note how the degree of $u$ has dropped again. 

We have $h'''(x)=\frac{12x^2+6}{\pi(1-x^2)^{5/2}}>0$, making $h'$ a strictly convex function. 
It is easy to verify by computer aid that $h'(-0.99)>0$, $h'(-0.66)<0$, and $h'(-0.33)>0$: thus the strictly convex function $h'$ has exactly two roots, one in each of the intervals $]-0.99,-0.66[$ and $]-0.66,-0.33[$. 
Via a numerical approximation we can find these roots $r_1\approx -0.859622$ and $r_2\approx -0.589783$. 
Thus $h(x)$ is strictly increasing on $]-1,r_1[$, decreasing on $]r_1,r_2[$, and increasing on $]r_1,0[$. 
Finally, $h(r_2)\approx 0.033>0$, thus $h$ indeed has a unique root $\alpha$, and it is in the interval $]-1,r_1[$. Hence, $\alpha<-0.5$. 
\end{proof}

\begin{rem}
The constant $0.87$ in the estimate $\Phi(x)\geq 0.87\cdot \Psi(x)$ is nearly optimal. 
Taylor expansions around $(0,0,0)$ show that the ratio $\Phi(x)/\Psi(x)$ converges to $\frac{9}{\pi^2}\approx 0.9119$ as $u_1,u_2,u_3$ tend to 0 simultaneously. 
Somewhat surprisingly, we cannot replace $0.87$ in the lemma by $\frac{9}{\pi^2}$. 
A similar analysis as in the proof shows that $(0,0,0)$ is in fact not the locus of the global minimum of the function $\Phi(x)-\frac{9}{\pi^2} \cdot \Psi(x)$: the minimum is at $(u,u,-2u)$ where $u\approx -0.67378$ with function value approximately $-0.0000377$. 
An elaborate computer analysis puts the optimal coefficient in the estimate around $0.908$; however this would require a rather meticulous calculation to verify. 
\end{rem}

We are ready to prove the main result of the paper. 

\begin{proof}
\emph{(of Theorem \ref{thm:main theorem})} First, let's investigate
what happens to the potential $\Phi$ at the moment of the potential interchange:
\[
\Phi(X_{\tau})=1-\frac{\mathbb{E}(\mathbf{T}(x))|_{x=X_{\tau}}}{\mu\frac{N^2}{27}}=1-\frac{27}{\mu N^{2}}\mathbb{E}\big(\mathbf{T}(X_{\tau})\,|\, X_{\tau}\big)=1-\frac{27}{\mu N^{2}}\mathbb{E}\big(\mathbf{T}-\tau\,|\, X_{\tau}\big)
\]
Hence, taking expectation yields
\[
\mathbb{E}\big(\Phi(X_{\tau})\big)=1-\frac{27}{\mu N^{2}}\mathbb{E}(\mathbf{T}-\tau)
\]
So now, we can estimate $\mathbb{E}(\mathbf{T})$ as:
\[
\mathbb{E}(\mathbf{T})=\mathbb{E}(\tau)+\mathbb{E}(\mathbf{T}-\tau)=\mathbb{E}(\tau)+\frac{\mu N^{2}}{27}\cdot\Big(1-\mathbb{E}\big(\Phi(X_{\tau})\big)\Big)\leq
\]
where we can apply Lemma \ref{lem:interchange of potentials}:
\[
\leq \mathbb{E}(\tau)+\frac{\mu N^{2}}{27}\cdot\Big(1-0.87\cdot \mathbb{E}\big(\Psi(X_{\tau})\big)\Big)=\mathbb{E}(\tau)+\frac{\mu N^{2}}{27}\cdot\Big(1-0.87\cdot \mathbb{E}(Y_{\tau})\Big)\leq
\]
So we can use the estimation of $Y_{\tau}$ proved in Corollary \ref{cor:recursion}:
\[
\leq \mathbb{E}(\tau)+\frac{\mu N^{2}}{27}\cdot\Big(1-0.87\cdot \big(4\varepsilon \mathbb{E}(\tau)+Y_{0}\big)\Big)=\frac{\mu N^{2}}{27}+\bigg(1-\frac{1.16}{9}\mu\varepsilon N^{2}\bigg)\mathbb{E}(\tau)-\frac{0.29}{9}\mu N^{2}Y_{0}
\]

To finish the proof, it suffices to show that the second term is negative, or equivalently, $1-\frac{1.16}{9}\mu\varepsilon N^{2}<0$. 
In the discrete version $\mu\varepsilon = 4\sin^2\left(\frac{\pi}{2N}\right)$, in the Poisson variant $\mu\varepsilon = 16\sin^2\left(\frac{\pi}{4N}\right)$, and in the Brownian version $\mu\varepsilon = \frac{\pi^2}{N^2}$. 
Thus $\mu\varepsilon N^2$ is roughly $\pi^2\approx 9.87$ in all three cases. 
To be more accurate, this is exactly the case for the Brownian version. 
In the other two variants, the worst constant is obtained for $N=3$: $4\sin^2\left(\frac{\pi}{2\cdot 3}\right)\cdot 3^2=9$ and $16\sin^2\left(\frac{\pi}{4\cdot 3}\right)\cdot 3^2\approx 9.65$, making the coefficient $1-\frac{1.16}{9}\mu\varepsilon N^{2}$ negative. 
To see the case of equality, note that in the last inequality we estimated from below $\mathbb{E}(\tau)$ by zero and $Y_{0}$ by zero as well. 
If we did not lose anything here, then $\mathbb{E}(\tau)=0$, thus we start from a three-token state. Moreover, $Y_{0}=0$ hence we started from the equidistant configuration by item 4. of Proposition~\ref{prop:properties of psi w/o proof}.
\end{proof}

\section{More general estimates}

We note that the last proof provides a somewhat stronger statement than the conjecture in all three variants. 
In the right hand side of the inequality $\mathbb{E}(\mathbf{T})\leq \frac{\mu N^{2}}{27}+\bigg(1-\frac{1.16}{9}\mu\varepsilon N^{2}\bigg)\mathbb{E}(\tau) -\frac{0.29}{9}\mu N^{2}Y_{0}$, 
the coefficient $1-\frac{1.16}{9}\mu\varepsilon N^{2}$ is at most $-0.23$ (assuming that $N\geq 5$, as otherwise $\mathbb{E}(\tau)=0$), so we obtain $\mathbb{E}(\mathbf{T}) + 0.23 \mathbb{E}(\tau)\leq \frac{\mu N^{2}}{27}-\frac{0.29}{9}\mu N^{2}Y_{0}$. 
In fact, Corollary~\ref{cor:recursion} can be slightly improved by noticing that the speed of the expected elevation of $\Psi$ is at least $2k\varepsilon\Delta t$ in states with $K=2\ell-1$ tokens. 
Putting $\tau_k:=\min\{t\ |\ K_{t}(x)=2\ell-1\}$, this yields the refined estimate 

\[
\mathbb{E}(\mathbf{T}) + 0.23 \mathbb{E}(\tau) + \sum\limits_{\ell=3}^{(\ell_0+1)/2} 0.61\mathbb{E}(\tau_\ell)\leq \frac{\mu N^{2}}{27}-\frac{0.29}{9}\mu N^{2}Y_{0}
\]

Informally, this means that even if we reward the process for being in states with more than three tokens by making the contribution of such a step a linear function of $\ell$ (roughly $0.61\ell$ rather than 1, as in the computation of the runtime $\mathbf T$), the maximum of the expected total contribution is still attained at the three-token equilibrium state.

We now focus on the discrete variant of the process, and show that the presented method can yield further estimates to the distribution of the runtime. 
We can view the process as an absorbing Markov chain; cf. \cite{KS76} for an introduction. 
As usual, the transition matrix is given in a canonical form: that is, indices corresponding to absorbing states (those with one token) are at the end, making the transition matrix a block matrix of the form $P=\begin{pmatrix} Q & R\\ 0 & I\end{pmatrix}$. 
Moreover, if the states are clustered according to the number of tokens in them, then $Q$ is also a block matrix, all of whose diagonal blocks are non-negative irreducible matrices. 
The spectral radius $\varrho$ of $Q$ carries an important probabilistic meaning: clearly, the supremum of those $\alpha\geq 1$ such that $\mathbb{E}(\alpha^\mathbf{T})$ is finite is $\varrho^{-1}$. 
Moreover, the vector $\underline{u'}$ of values $\mathbb{E}(\alpha^\mathbf{T})$ assigned to all non-absorbing states is the restriction of the unique solution to the system of linear equations $\begin{pmatrix} \alpha Q & \alpha R\\ 0 & I\end{pmatrix} \underline{u} =  \underline{u}$, where the coordinates of $ \underline{u}$ corresponding to absorbing states are all 1. 
Equivalently, $\begin{pmatrix} Q & R\\ 0 & I\end{pmatrix} \underline{u} =  \underline{v}$, where components of $\underline{v}$ corresponding to absorbing states are still 1, and the rest is filled with $1/\alpha$ times the values $\mathbb{E}(\alpha^\mathbf{T})$, that is $\underline{v}'=(1/\alpha)\underline{u}'$. 
As we are interested in such expected values, we compute the spectral radius $\varrho$ of $Q$, and provide a formula to $\mathbb{E}(\alpha^\mathbf{T})$ for three-token initial states. 

\begin{lem}\label{lem:geomquot}
Given the discrete version of the protocol with parameters $p$, $N$ and $\varepsilon=4p(1-p)\sin^{2}\big(\frac{\pi}{2N}\big)$. 
Then the spectral radius of $Q$ is $\varrho=1-4p(1-p)\sin^2\left(\frac{\pi}{N}\right)$. 
In particular, $\varrho\approx 1-4\varepsilon$, and we have the precise bounds $1-4\varepsilon\leq \varrho\leq 1-3\varepsilon$. 
Moreover, given an $\alpha\geq 1$, the expected value $\mathbb{E}(\alpha^{\mathbf{T}})$ is finite if and only if $\alpha<\varrho^{-1}$, and then for three-token states with distances $a,b,c$ between the tokens it is $\mathbb{E}(\alpha^\mathbf{T}) = \frac{\beta^a-\beta^{N-a}+\beta^b-\beta^{N-b}+\beta^c-\beta^{N-c}}{1-\beta^N}$, where $\beta$ is any of the two solutions of the equation $\beta+\beta^{-1}=\frac{\alpha^{-1}-1}{p(1-p)}+2$. 
\end{lem}
\begin{proof}
According to the min-max Collatz-Wielandt formula \cite{C42}, the spectral radius $\varrho$ is bounded from above by $\max\limits_j \frac{[Q\underline{x}]_j}{x_j}$ for any positive vector $\underline{x}$ with j-th coordinate $x_j$. 
Let $\underline{1}$, $\underline{\psi}$ and $\underline{k}$ be the vectors that assign to any non-absorbing state the value $1$, $\Psi$ and $K$, respectively. 
Note that both $\underline{1}-\underline{\psi}$ and $\underline{k}-\underline{\psi}$ are zero in the absorbing states. 
Thus the evolution of $\Psi$, that is, Lemma~\ref{lem:recursion w proof} translates to the equation $Q(\underline{1}-\underline{\psi}) = (\underline{1}-\underline{\psi})-\varepsilon(\underline{k}-\underline{\psi})$. 
At a given non-absorbing state with index $j$, potential value $\Psi$ and number of tokens $K$, we have 
\[\frac{[Q(\underline{1}-\underline{\psi})]_j}{(\underline{1}-\underline{\psi})_j}= \frac{((\underline{1}-\underline{\psi})-\varepsilon(\underline{k}-\underline{\psi}))_j}{(\underline{1}-\underline{\psi})_j}= 1-\varepsilon\frac{K-1}{1-\Psi}-\varepsilon\frac{1-\Psi}{1-\Psi} \leq 1 - \varepsilon\frac{3-1}{1-\Psi} -\varepsilon\leq 1-\varepsilon -\varepsilon\frac{3-1}{1-0}=1-3\varepsilon
\]
yielding $\varrho\leq 1-3\varepsilon$. 
We can repeat the same argument for the truncated absorbing chain that halts when a state with at most three tokens is reached. 
Let $Q'$ be the corresponding upper-left corner: that is, the upper-left square of the transition matrix corresponding to states with at least five tokens. 
In that case, it is not possible to turn the formula for the evolution of $\Psi$ to a clear matrix equation that only involves $Q'$: the problem is that the vector $(\underline{1}-\underline{\psi})$ has non-zero coordinates in the new absorbing states (those with three tokens). 
However, as those values are positive, we still have the inequality $Q'(\underline{1'}-\underline{\psi'}) \leq (\underline{1'}-\underline{\psi'})-\varepsilon(\underline{k'}-\underline{\psi'})$, where the primes represent restriction to states with at least five tokens. 
By repeating the above calculation, we obtain 
\[\frac{[Q'(\underline{1'}-\underline{\psi'})]_j}{(\underline{1'}-\underline{\psi'})_j}\leq \frac{((\underline{1'}-\underline{\psi'})-\varepsilon(\underline{k'}-\underline{\psi'}))_j}{(\underline{1'}-\underline{\psi'})_j}= 1-\varepsilon\frac{K-1}{1-\Psi}-\varepsilon\frac{1-\Psi}{1-\Psi} \leq 1 -\varepsilon\frac{5-1}{1-\Psi}-\varepsilon\leq 1-\varepsilon -\varepsilon\frac{5-1}{1-0}=1-5\varepsilon
\]
So all the blocks corresponding to states with at least five tokens have a spectral radius at most $1-5\varepsilon$. 
Hence, in order to prove that the spectral radius of $Q$ is indeed $\varrho=1-4p(1-p)\sin^2\left(\frac{\pi}{N}\right)=1-4p(1-p)4\sin^2\left(\frac{\pi}{2N}\right)\cos^2\left(\frac{\pi}{2N}\right)\geq 1-4\varepsilon$, we need to show the same assertion to the block corresponding to three-token states. 
That is, we are only interested in the process starting from three-token initial states. 

Let $g(a,b,c) = \frac{\beta^a-\beta^{N-a}+\beta^b-\beta^{N-b}+\beta^c-\beta^{N-c}}{1-\beta^N}$ and let $1\leq \alpha<\varrho^{-1}$. 
If $f(a,b,c)$ denotes $\mathbb{E}\left(\alpha^{\mathbf{T}}\right)$ from the initial state with distances $a,b,c$, then the linear equation obtained from first step analysis (see the explanation before the lemma) is  
\[
\alpha^{-1}f(a,b,c)=(p^3+(1-p)^3)f(a,b,c) + p^2(1-p)(f(a-1,b+1,c)+f(a,b-1,c+1)+f(a+1,b,c-1))+
\]
\[
+p(1-p)^2(f(a+1,b-1,c)+f(a,b+1,c-1)+f(a-1,b,c+1))
\]
If $(a,b,c)$ represents an absorbing state (with one token), that is, $a,b$ or $c$ is 0, then $f(a,b,c)=1$. 
(We only write the equations for non-absorbing triples $(a,b,c)$ on the left, but an absorbing triple can occur on the right hand side.) 
Clearly $\mathbb{E}(\alpha^\mathbf{T})$ is continuous in the given range of $\alpha$, and the expression $g(a,b,c)$ claimed to be equal to it is also continuous between singularities. 
The smallest singularity of $g(a,b,c)$ for $\alpha\in [1,\infty[$ appears when $\beta^N=1$ and $\beta+\beta^{-1}$ is largest possible, that is, when $\beta=e^{\frac{2\pi}{N}i}$. 
Then $\beta+\beta^{-1} = 2\cos\left(\frac{2\pi}{N}\right)= 2 - 4\sin^{2}\left(\frac{\pi}{N}\right)$, making $\alpha^{-1} = 1 - 4p(1-p)\sin^{2}\left(\frac{\pi}{N}\right)$. 
Thus in the range $1\leq\alpha<(1 - 4p(1-p)\sin^{2}\left(\frac{\pi}{N}\right))^{-1}$, the expression $g(a,b,c)$ is indeed finite (and continuous). 
Also note that $g(a,b,c)=1$ for absorbing states. 

We show that $g$ is a solution to all the equations above. 
This can be verified by a straightforward calculation 
\[
(p^3+(1-p)^3)g(a,b,c) + p^2(1-p)(g(a-1,b+1,c)+g(a,b-1,c+1)+g(a+1,b,c-1))+
\]
\[
+p(1-p)^2(g(a+1,b-1,c)+g(a,b+1,c-1)+g(a-1,b,c+1)) = 
\]
\[
(p^3+(1-p)^3)g(a,b,c) +   \frac{\beta^a+\beta^b+\beta^c}{1-\beta^N}\left(p^2(1-p)\left(\beta^{-1}+1+\beta\right)+p(1-p)^2\left(\beta+1+\beta^{-1}\right)\right) - 
\]
\[
- \frac{\beta^{N-a}+\beta^{N-b}+\beta^{N-c}}{1-\beta^N}\left(p^2(1-p)\left(\beta+1+\beta^{-1}\right)+p(1-p)^2\left(\beta^{-1}+1+\beta\right)\right) = 
\]
\[
g(a,b,c)\left(p^3+(1-p)^3+\left(p^2(1-p)+p(1-p)^2\right)\left(\beta+1+\beta^{-1}\right)\right) =
\]
\[
= g(a,b,c)\left(1+p(1-p)\left(\beta+\beta^{-1}-2\right)\right) = g(a,b,c)\alpha^{-1}
\]

To summarize: given any $1\leq\alpha<(1 - 4p(1-p)\sin^{2}\left(\frac{\pi}{N}\right))^{-1}$ the expressions $g(a,b,c)$ form a vector that is all 1 in absorbing states, and that satisfies the matrix equation $\begin{pmatrix} \alpha Q & \alpha R\\ 0 & I\end{pmatrix} \underline{u} =  \underline{u}$. 
Hence, in the given range $1\leq\alpha<(1 - 4p(1-p)\sin^{2}\left(\frac{\pi}{N}\right))^{-1}$, we have $\mathbb{E}(\alpha^{\mathbf{T}})=g$. 
As $\alpha\rightarrow \left(1 - 4p(1-p)\sin^{2}\left(\frac{\pi}{N}\right)\right)^{-1}$, the formula $g(a,b,c)$ tends to infinity, thus the expected values cannot be finite for $\alpha = \left(1 - 4p(1-p)\sin^{2}\left(\frac{\pi}{N}\right)\right)^{-1}$. 
Hence, $\varrho = 1 - 4p(1-p)\sin^{2}\left(\frac{\pi}{N}\right)$. 
\end{proof}

\begin{lem}\label{lem:stopexp}
Given the discrete version of the protocol with parameters $p$, $N$ and $\varepsilon=4p(1-p)\sin^{2}\big(\frac{\pi}{2N}\big)$. 
Let $1\leq \alpha\leq (1-\varepsilon)^{-1}$. 
Then 
\begin{itemize}
\item $\varepsilon\alpha \mathbb{E}\left(\frac{\alpha^{\mathbf{\tau}}-1}{\alpha-1}(5-Y_\tau)\right)\leq \mathbb{E}(Y_\tau)-Y_0$, and 
\item $2\varepsilon\alpha \mathbb{E}\left(\frac{\alpha^{\mathbf{\mathbf{T}}}-1}{\alpha-1}\right)\leq 1-Y_0$.
\end{itemize}
\end{lem}
\begin{proof}
We focus on the first inequality; the second one can be shown analogously. 
The starting point of the calculation is a slightly rephrased form of Lemma~\ref{lem:recursion w proof}, namely $\mathbb{E}(Y_{t+1}\mid\mathcal{F}_{t})=(1-\varepsilon)Y_t+\varepsilon K_{t}$. 
Multiplying both sides by $\alpha^{t+1}$ results in  
\[
\mathbb{E}(\alpha^{t+1}Y_{t+1}\mid\mathcal{F}_{t})=(1-\varepsilon)\alpha^{t+1}Y_t+\varepsilon \alpha^{t+1}K_{t}
\]

If $0\leq t< \tau$, then $K_t\geq 5$, which yields 
\[
\mathbb{E}(\alpha^{t+1}Y_{t+1}-\alpha^{t}Y_t\mid\mathcal{F}_{t}) \geq ((1-\varepsilon)\alpha^{t+1}-\alpha^t)Y_t+5\varepsilon \alpha^{t+1} = ((1-\varepsilon)\alpha-1)\alpha^tY_t+5\varepsilon\alpha \alpha^{t}
\]

Let us apply the operator $\mathbb{E}(\sum\limits_{t=0}^{\tau-1} .)$ to both ends. 
Every summand contributes a convergent sum by Lebesgue's Dominated Convergence, as $1\leq \alpha\leq (1-\varepsilon)^{-1}$ is well below $\varrho^{-1}$. 
The telescoping sum on the left yields $\mathbb{E}(\alpha^{\tau}Y_{\tau}-Y_0)$. 
Note that the constant $((1-\varepsilon)\alpha-1)$ is at most 0, and $Y_t$ is increasing in expectation according to the above evolution formula. 
That is, $\mathbb{E}(Y_\tau)\geq \mathbb{E}(Y_t)$ for all $t<\tau$. 
We obtain 

\[
\mathbb{E}(\alpha^{\tau}Y_{\tau}-Y_0) \geq \sum\limits_{t=0}^{\tau-1} \mathbb{E}(((1-\varepsilon)\alpha-1)\alpha^tY_t)+5\varepsilon\alpha \sum\limits_{t=0}^{\tau-1} \mathbb{E}(\alpha^{t}) \geq 
((1-\varepsilon)\alpha-1)\mathbb{E}\left(\frac{\alpha^\tau-1}{\alpha-1}Y_\tau\right)+5\varepsilon\alpha \mathbb{E}\left(\frac{\alpha^\tau-1}{\alpha-1}\right)\geq
\]
\[
\geq \varepsilon\alpha \mathbb{E}\left(\frac{\alpha^\tau-1}{\alpha-1}(5-Y_\tau)\right) + \mathbb{E}\left((\alpha^\tau-1)Y_\tau\right)
\]

By subtracting $\mathbb{E}\left((\alpha^\tau-1)Y_\tau\right)$, we have 
\[
\mathbb{E}(Y_\tau-Y_0) \geq \varepsilon\alpha \mathbb{E}\left(\frac{\alpha^\tau-1}{\alpha-1}(5-Y_\tau)\right)
\]
as claimed. 

The other item can be shown similarly: if we switch the stopping time $\tau$ to $\mathbf T$, the constant 5 has to be replaced by 3, since we can only guarantee $K_t\geq 3$ if $t<\mathbf T$. 
This results in the inequality $\varepsilon\alpha \mathbb{E}\left(\frac{\alpha^{\mathbf{T}}-1}{\alpha-1}(3-Y_\mathbf{T})\right)\leq \mathbb{E}(Y_\mathbf{T})-Y_0$, where $Y_\mathbf{T}=1$. 
\end{proof}

We note that the second item provides a quadratic upper bound for $\mathbb{E}(\mathbf{T})$ by putting $\alpha\rightarrow 1$. 
Indeed, the left hand side converges to $2\varepsilon \mathbb{E}\left(\mathbf{T}\right)$ as $\alpha\rightarrow 1$, thus $2\varepsilon \mathbb{E}\left(\mathbf{T}\right)\leq 1$, and consequently $\mathbb{E}\left(\mathbf{T}\right)\leq \frac{1}{2\varepsilon} \approx \frac{2N^2}{\pi^2}\approx 0.203 N^2$. 
This bound is never tight, as we demonstrated in Theorem \ref{thm:main theorem}: the tight bound is $\frac{4}{27}N^2\approx 0.148 N^2$. 
This is the reason we had to cut the process in two: we first estimate the parameters until a three-token state is reached, and then use the precise formulas to the parameters for three-token initial states. 
However, for one particular choice of $\alpha$, the second item of the above lemma can yield a tight bound. 

\begin{cor}\label{cor:specexp}
Given the discrete version of the protocol with parameters $p$, $N$ and $\varepsilon=4p(1-p)\sin^{2}\big(\frac{\pi}{2N}\big)$, we have 
\[
\mathbb{E}\bigg(\Big(\frac{1}{1-\varepsilon}\Big)^{\mathbf{T}}\bigg)\leq\frac{3}{2}
\]
 with equality if and only if we start from the equidistant three-token
configuration.\label{thm:weaker theorem}
\end{cor}
Note that this statement provides a tight bound to a linear combination of the $\mathbb{P}(\mathbf{T}\ge t)$'s
with the weights $(1-\varepsilon)^{-t}-(1-\varepsilon)^{-(t-1)}=\varepsilon(1-\varepsilon)^{-t}$.
\begin{proof}
By applying the second item of Lemma~\ref{lem:stopexp} for $\alpha=(1-\varepsilon)^{-1}$, we have 
\[
1\geq 1-Y_0\geq 2\varepsilon(1-\varepsilon)^{-1} \mathbb{E}\left(\frac{(1-\varepsilon)^{-\mathbf{\mathbf{T}}}-1}{(1-\varepsilon)^{-1}-1}\right) = 
2\varepsilon \mathbb{E}\left(\frac{(1-\varepsilon)^{-\mathbf{\mathbf{T}}}-1}{1-(1-\varepsilon)}\right) = 
2\mathbb{E}\left((1-\varepsilon)^{-\mathbf{\mathbf{T}}}-1\right)
\]

The case of equality holds exactly if we did not lose anything in the estimations. 
Those in Lemma~\ref{lem:stopexp} that were used in the second item are tight if and only if $K_0=3$. 
(Note that the constant $(1-\varepsilon)\alpha-1=0$ if $\alpha=(1-\varepsilon)^{-1}$.)
Equality in $1\geq 1-Y_0$ holds if and only if $Y_0=0$, which is equivalent to the assumption that the tokens are distributed equidistantly by item 4. of Lemma~\ref{prop:properties of psi w proof}.
\end{proof}

\begin{cor}\label{cor:spectelescope}
Given the discrete version of the protocol with parameters $p$, $N$ and $\varepsilon=4p(1-p)\sin^{2}\big(\frac{\pi}{2N}\big)$, we have 
\[
\mathbb{E}\left(\frac{5-Y_\tau}{(1-\varepsilon)^{\tau}}\right)\leq\ 5
\]
 with equality if and only if we start from the equidistant three-token
configuration.\label{thm:weaker theorem}
\end{cor}

\begin{proof}
By applying the first item of Lemma~\ref{lem:stopexp} for $\alpha=(1-\varepsilon)^{-1}$, we have 
\[
\mathbb{E}(Y_\tau)\geq \mathbb{E}(Y_\tau)-Y_0 \geq \varepsilon(1-\varepsilon)^{-1} \mathbb{E}\left(\frac{(1-\varepsilon)^{-\mathbf{\tau}}-1}{(1-\varepsilon)^{-1}-1}(5-Y_\tau)\right)  = \varepsilon(1-\varepsilon)^{-1} \mathbb{E}\left(\frac{(1-\varepsilon)^{-\mathbf{\tau}}-1}{\varepsilon(1-\varepsilon)^{-1}}(5-Y_\tau)\right)= 
\]
\[
\mathbb{E}\left(((1-\varepsilon)^{-\mathbf{\tau}}-1)(5-Y_\tau)\right) = \mathbb{E}\left(\frac{5-Y_\tau}{(1-\varepsilon)^{\tau}}\right) - 5 + \mathbb{E}(Y_\tau)
\]
\end{proof}

\begin{thm}\label{thm:genint}
Given the discrete version of the protocol with parameters $p$, $N$ and $\varepsilon=4p(1-p)\sin^{2}\big(\frac{\pi}{2N}\big)$. 
Let $1\leq \alpha< (1-\varepsilon)^{-1}$, and let $\gamma=-\log_{1-\varepsilon} \alpha$. 
For a three-token state $x$, let $g(x)$ be the expected value of $\alpha^{\mathbf{T}}$ with initial position $x$; cf. Lemma~\ref{lem:geomquot} for the precise formula. 

Then $\sup\limits_{K(x)=3} \frac{g(x)}{(1-\Psi(x)/5)^\gamma}$ is an upper estimate for $\mathbb{E}\left(\alpha^{\mathbf{T}}\right)$ with arbitrary initial state. 
In particular, if the function $\frac{g(x)}{(1-\Psi(x)/5)^\gamma}$ defined on all three-token states attains its maximum at the three-token equidistant state, then so does $\mathbb{E}\left(\alpha^{\mathbf{T}}\right)$. 
\end{thm}

We remark that by plotting the function $\frac{g(x)}{(1-\Psi(x)/5)^\gamma}$ it seems evident that the maximum is indeed attained at the three-token equidistant state. 
However, we do not have a rigorous verification similar to the proof Lemma~\ref{lem:interchange of potentials}, due to the fact that the current function is more complicated to analyze. 

\begin{proof}
Given any initial state, consider the probability distribution of the set of pairs $\{(x,t) \mid K(x)=3, t\in \mathbb{N}_0\}$ induced by the process: namely, the probability that $t=\tau$ and $x_\tau=x$, that is, we hit the set of three-token states at time $t$ and in the particular state $x$. 
Then 
\[
\mathbb{E}\left(\alpha^{\mathbf{T}}\right) = \mathbb{E}(\alpha^tg(x)) \leq \left(\sup\limits_{K(x)=3} \frac{g(x)}{(5-\Psi(x))^\gamma}\right)\cdot \mathbb{E}\left(\alpha^t(5-\Psi(x))^\gamma\right) = \left(\sup\limits_{K(x)=3} \frac{g(x)}{(5-\Psi(x))^\gamma}\right)\cdot \mathbb{E}\left(\left(\frac{5-\Psi(x)}{(1-\varepsilon)^t}\right)^\gamma\right)
\]

By using Jensen's inequality here, and then later Corollary~\ref{cor:spectelescope}, we obtain 

\[
\mathbb{E}\left(\alpha^{\mathbf{T}}\right) \leq \left(\sup\limits_{K(x)=3} \frac{g(x)}{(5-\Psi(x))^\gamma}\right)\cdot \mathbb{E}\left(\frac{5-\Psi(x)}{(1-\varepsilon)^t}\right)^\gamma \leq \left(\sup\limits_{K(x)=3} \frac{g(x)}{(5-\Psi(x))^\gamma}\right)\cdot 5^\gamma = \sup\limits_{K(x)=3} \frac{g(x)}{(1-\Psi(x)/5)^\gamma}
\]

As for the second assertion of the theorem, if the maximum is attained at the three-token equidistant state $x_0$, then in that state we have $\Psi(x_0)=0$, thus $\frac{g(x_0)}{(1-\Psi(x_0)/5)^\gamma}= g(x_0)$ is exactly the expected value of $\mathbb{E}\left(\alpha^{\mathbf{T}}\right)$ with the three-token equidistant state as the initial state. 
\end{proof}

\section{Closing remarks and open problems}

%In this paper, we have provided a proof to the Herman Protocol Conjecture independently from \cite{BGKOW15}, and generalized the result in several directions. 
The question whether the probabilities $\mathbb{P}(\mathbf{T}\geq t)$ are maximal for all $t$ in the three-token equidistant configuration remains open, and it seems out of reach at the moment. 
It would be interesting to extend Theorem~\ref{thm:genint} to the whole interval $[1,1/\varrho[$ where the expectation is finite; cf. Lemma~\ref{lem:geomquot}. 
Note that this certainly requires some new ideas: in the proof of Theorem~\ref{thm:genint} we use Jensen's inequality for the real concave function $x^\gamma$ where $0<\gamma\leq 1$. 
If $\gamma>1$, then the inequality is reversed, as $x^\gamma$ is convex rather than concave. 
A similar problem occurs when trying to extend the domain to the left, say, to $[0,1[$. 

It also seems plausible to show that every moment of $\mathbf{T}$ attains its maximum on the three-token state (in all three variants of the protocol). 

Another natural problem is to study $\mathbb{E}(\mathbf{T})$ for initial states where there is a token in every original position, i.e., the essentially unique equidistant $N$-token state (for odd $N$). Surprisingly, it is useful to combine the two completely different methods in the present paper and in \cite{BGKOW15}. 

\begin{prop}
Given an odd integer $N\geq 3$ and $p=1/2$. 
Let $\mathbf{T}$ be the runtime of the (unbiased) discrete version of the process from the equidistant $N$-token state. 
Then for large enough $N$ we have $CN^2<\mathbb{E}(\mathbf{T})$, where $C\approx 0.072$.
\end{prop}
\begin{proof}
The vector of expected times to absorption (with coordinates corresponding to the transient states as initial states) in a finite absorbing Markov chain is $Z\underline{1}$, where $Z=(I-Q)^{-1}$ is the fundamental matrix of the chain and  $\underline{1}$ is the all-one vector; cf. \cite{KS76}. 
More generally speaking, $Z\underline{v}$ is the expected sum of the entries of the vector $\underline{v}$ corresponding to the states reached during the walk. 
The equation shown in the proof of Lemma~\ref{lem:geomquot}, that is, 
$Q(\underline{1}-\underline{\psi}) = (\underline{1}-\underline{\psi})-\varepsilon(\underline{k}-\underline{\psi})$, can be translated to $Z(\underline{k}-\underline{\psi})=\frac{1}{\varepsilon}(\underline{1}-\underline{\psi})$. 
In \cite{BGKOW15}, one of the crucial lemmas to prove the biased discrete variant of the Herman Protocol Conjecture was Lemma~3, which was first shown in \cite{FZ15}. 
In our terminology, it translates to $(I-Q)\underline{v_3}=\frac{1}{2}(\underline{k}-\underline{1})$, that is, $Z(\underline{k}-\underline{1})= 2\underline{v_3}$. 
Here, $\underline{v_3}$ is the vector whose coordinate in a transient state $x$ is $4N^2f_3^{(K)}(h(x)/N)$, where $f_3^{(K)}$ is the cubic polynomial $f_3^{(K)}({\bf{y}})=\sum y_{i_1}y_{i_2}y_{i_3}$ with the summation running through those triples of indices $1\leq i_1<i_2<i_3\leq K$ such that both $i_2-i_1$ and $i_3-i_2$ are odd, and $h(x)$ is the gap vector containing the list of distances between the $K$ consecutive pairs of tokens in the given state $x$. 

By combining the above two equations involving $Z$, we obtain $\frac{1}{\varepsilon}(\underline{1}-\underline{\psi})-2\underline{v_3}=Z(\underline{1}-\underline{\psi})\leq Z\underline{1}=\mathbb{E}(\mathbf{T})$. 
In the $N$-token equidistant state, the entry of $\underline{1}-\underline{\psi}$ is 1 according to item 4. of Proposition~\ref{prop:properties of psi w proof}, the entry of $\underline{v_3}$ is approximately $\frac{N^2}{6}$, and $\frac{1}{\varepsilon}\approx \frac{4N^2}{\pi^2}$, yielding the lower estimate for $\mathbb{E}(\mathbf{T})$. 
\end{proof}

It is reasonable to assume that $\mathbf{T}_N/N^2$ is convergent in distribution; here we emphasize that the runtime $\mathbf{T}_N$ depends on $N$, and only on $N$, since the initial position is fixed. Intuitively, the three variants of the protocol evolve similarly. 
Hence, by the scale invariance of the Wiener process, if the perimeter of the circle is 1, and the distance between two tokens is the actual (half-)length of the shorter arc connecting them, then spreading more and more tokens uniformly on a circle, and moving each of them via i.i.d. Brownian motions with variance 1, the process should fizzle out in a time with finite expectation $\lim\limits_{N\rightarrow \infty} \mathbb{E}(\mathbf{T}_N/N^2)$ and limiting distribution that is the weak limit of $\mathbf{T}_N/N^2$. 

This can provide a simple viewpoint to the limiting process where one-dimensional Brownian motions are uniformly spread all over a circle. 
For the analogous game on a line rather than a circle, see the recent paper \cite{HOV21}.

In fact, it is possible that the three variants are even more closely related. 
We can compute $\mathbb{E}\left(\alpha^{\mathbf{T}}\right)$ for three-token initial states in all three variants, yielding similar formulas as the one in Lemma~\ref{lem:geomquot}. 
In the Poisson variant with arbitrary $p$, we have $\mathbb{E}\left(\alpha^{\mathbf{T}}\right)= \frac{\beta^a-\beta^{N-a}+\beta^b-\beta^{N-b}+\beta^c-\beta^{N-c}}{1-\beta^N}$ where $\beta$ is any of the two solutions of the equation $\beta + \beta^{-1}= 2- \log \alpha$. 
Similarly, in the Brownian version, we have $\mathbb{E}\left(\alpha^{\mathbf{T}}\right)= \frac{\beta^a-\beta^{N-a}+\beta^b-\beta^{N-b}+\beta^c-\beta^{N-c}}{1-\beta^N}$ where $\beta$ is any of the two solutions of the equation $\log^2 \beta = 4\log \alpha$. 

The similarity of these three formulas suggests that there might be a correspondence between (a pair of) the three versions. 
Perhaps each pair $(N, \alpha)$ corresponds to some $(N', \alpha')$ such that $\mathbb{E}\left(\alpha^{\mathbf{T}}\right)$ for the Poisson variant with parameters $p=1/2, N$ coincides with $\mathbb{E}\left(\alpha'^{\mathbf{T}}\right)$ for the discrete variant with parameters $p=1/2, N'$ (for all possible initial states). 
A similar question can be posed involving any pair of the three variants. 
In principle, even the distribution of $\alpha^{\mathbf{T}}$ could coincide after a proper identification; this, however is only plausible for an identification between the discrete and the Poisson variants, as the range of $\alpha^{\mathbf{T}}$ is countable in these two versions, but not in the third one. 

Should any such correspondence exist, it would have to be complicated. 
By applying the scale invariance of the Wiener process once again, it is clear that in the Brownian version the pair of parameters $(N, \alpha)$ can be replaced by just the single parameter $\alpha^{-N^2}$. If the expression $\alpha^{-N^2}$ coincide for two pairs $(N_1, \alpha_1)$ and $(N_2, \alpha_2)$, then $\mathbb{E}\left(\alpha^{\mathbf{T}}\right)$ is the same for any initial state, where the initial state is represented by the ratio of all (half) arc lengths over the perimeter of the circle for each arc connecting consecutive tokens. 
For example, for three-token states, we consider two initial states with the two different pairs of parameters $(N_1, \alpha_1)$ and $(N_2, \alpha_2)$ equivalent, if $(a/N, b/N, c/N)$ yields the same triple (up to a cyclic permutation). 

Using computer assistance we have verified that no such phenomenon occurs neither in the discrete nor in the Poisson variant. 

In the discrete version, we chose $N=5$ and $\alpha=1.083287067675$. 
The only 5-token state yielded the expected value $\mathbb{E}\left(\alpha^{\mathbf{T}}\right)=1.28979128$. For $N=10$ and $\alpha=1.019997333759931$, which is the $\alpha$ value we need to pick so that the expectations coincide for all three-token states (cf. the formula for $\mathbb{E}\left(\alpha^{\mathbf{T}}\right)$ in the discrete variant), we obtained $\mathbb{E}\left(\alpha^{\mathbf{T}}\right)=1.29889246$ for the equidistant five-token state. 

In the Poisson version, we chose $N=5$ and $\alpha=1.083287067675$. 
The only 5-token state yielded the expected value $\mathbb{E}\left(\alpha^{\mathbf{T}}\right)=1.0677968$. For $N=10$ and $\alpha=1.02030439850089$, which is the $\alpha$ value we need to pick so that the expectations coincide for all three-token states (cf. the formula for $\mathbb{E}\left(\alpha^{\mathbf{T}}\right)$ in the Poisson variant), we obtained $\mathbb{E}\left(\alpha^{\mathbf{T}}\right)=1.06577509$ for the equidistant five-token state. 

Hence, these numerical results suggest that there shouldn't be a reasonably simple correspondence between any one of these two versions and the Wiener variant. 
We pose it as a question here to turn this anticipation into a rigorous proof, and to clarify the analogous question for the two discrete-time variants as well. 
We expect a negative answer, which justifies the separate treatment of the three versions throughout this paper.

\bibliographystyle{plain}
\bibliography{refs}

\end{document}